# Switching magnon chirality in artificial antiferromagnet

Yahui Liu[1], Zhengmeng Xu[1], Lin Liu[1], Kai Zhang[1], Yang Meng[2,3,*], Yuanwei Sun[1,4], Peng Gao[1,4], Hong-Wu Zhao[2,3,5], Qian Niu[6,†] & J. Li[1,‡]

[1]International Center for Quantum Materials, School of Physics, Peking University, Beijing 100871, China
[2]Beijing National Laboratory for Condensed Matter Physics, Institute of Physics, Chinese Academy of Sciences, Beijing 100190, China
[3]School of Physical Sciences, University of Chinese Academy of Sciences, Beijing 100049, China
[4]Electron Microscopy Laboratory, School of Physics, Peking University, Beijing 100871, China
[5]Songshan Lake Materials Laboratory, Dongguan 523808, China
[6]Department of Physics, University of Texas at Austin, Austin, Texas 78712, USA

* Corresponding author: ymeng@iphy.ac.cn
† Corresponding author: niu@physics.utexas.edu
‡ Corresponding author: jiali83@pku.edu.cn

Magnons in antiferromagnets can support both right-handed and left-handed chiralities, which shed a light on the chirality-based spintronics. Here we demonstrate the switching and reading of magnon chirality in an artificial antiferromagnet. The coexisting antiferromagnetic and ferromagnetic characteristic resonance modes are discovered, which permits a high tunability in the modulation of magnon chirality. The reading of the chirality is accomplished via the chirality-dependent spin pumping as well as spin rectification effect. Our result illustrates an ideal antiferromagnetic platform for handling magnon chirality and paves the way for chirality-based spintronics.

In condensed matter physics, the chirality of elementary particles and quasiparticles plays an important role in many unconventional phenomena, such as the quantum Hall effect [1] and chiral phonon excitations [2]. Meanwhile, chirality provides an intrinsic degree of freedom, which may be utilized as information carriers in the transmission and processing of information (e.g., polarization of photons) [3]. In magnetic materials, magnons (quantized spin excitations) are chiral quasiparticles while the chirality of magnons has received little interest because ferromagnets (FMs) only support right-handed magnon chirality. Recently the magnons in antiferromagnets (AFMs) [4,5,6] were proposed to carry both right-handed and left-handed chiralities [7,8]. Due to the crucial advantages of magnetic materials such as the compatibility with existing technology and ease of nanopatterning, chiral magnons are the prospective information carriers with high fidelity and low-energy cost. Hence, it is urgent to find a magnonic platform with broad tunability where the magnon chirality could be easily manipulated and definitely measured.

In this work, we demonstrate the switching and reading of magnon chirality in an artificial AFM, which succeeds in combining the unique merits of FMs and AFMs. The magnetic resonance modes of FM and AFM characteristics both exist [Fig. 1(c) to (e)], providing the high adjustability and easy control of magnon chirality. Meanwhile, by spatially separating two magnetic sublattices of the artificial AFM, we can selectively probe the magnetic contribution to spin pumping from one specific magnetic sublattice. Our innovative design enables the electrical readout of the right-handed and left-handed chiralities via spin pumping in a quantitative manner [Fig. 1 (a) and (b)], as well as via spin rectification effect. Furthermore, the chirality-dependent spin pumping of FM and AFM characteristic resonance modes were discovered in coexistence, permitting the modulation of the chiralities [Fig. 1(f)], which is vital for the applications of chiral magnons as information carriers.

An artificial AFM consists of two magnetic sublattices, $M_A$ and $M_B$, with AFM coupling. When these two magnetic sublattices are uncompensated and in perfectly antiparallel alignment, the net moment ($|M_A - M_B|$) behaves as a ferromagnet. In ferromagnetic resonance (FMR), the net moment precesses around an external magnetic field $H$ with right-handed chirality, leading to right-handed chirality of the greater moment (master) between $M_A$ and $M_B$ and left-handed chirality of the weaker moment (slave) around their equilibrium directions. We can reach the $M_A$ master ($M_A > M_B$) or $M_B$ master ($M_A < M_B$) phase by resizing $M_A$ and $M_B$ in the artificial AFM, resulting in the right-handed or left-handed chirality of $M_A$ precession about the +z direction [Fig. 1(c) and (d)]. This precession mode will be referred to as the FMR mode. In the $M_B$ master phase, a sufficiently strong magnetic field may twist $M_A$ and $M_B$ into a canted state. The AFM coupling of $M_A$ and $M_B$ leads to the right-handed chirality of $M_A$ precession about the +z direction [Fig. 1(e)], which will be referred to as the exchange mode [9,10] in the following discussion. This exchange mode is the characteristic mode of AFM resonance, in comparison to the FMR mode of FM feature. Hence, the chirality of $M_A$ precession can be manipulated by resizing or twisting $M_A$ and $M_B$.

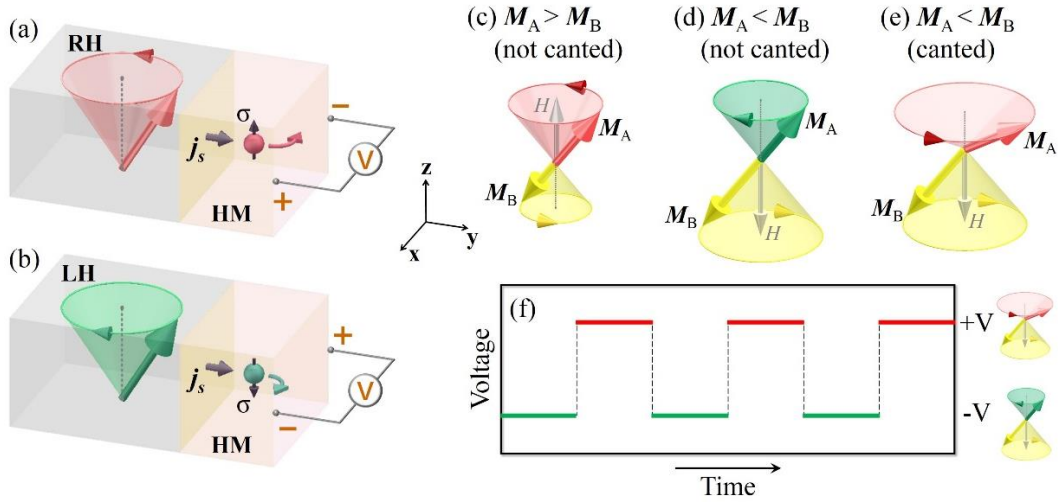

Fig. 1. Schematic of chirality-dependent spin pumping. Magnetization precessions with (a) right-handed (RH) and (b) left-handed (LH) chiralities around the equilibrium axis (+z axis) pump spin current into a heavy metal (HM) layer. The DC voltage polarity detects the spin polarization (+z or -z) of the spin current. (c) to (e) sketch the magnetization precessions of two magnetic sublattices $M_A$ and $M_B$ with AFM coupling. When $M_A$ and $M_B$ are collinearly aligned (FM mode), as shown in (c) and (d), magnon chirality can be switched between RH and LH by resizing $M_A$ and $M_B$. LH chirality of $M_A$ precession in (d) can be switched to RH chirality in (e) by twisting $M_A$ and $M_B$ into a canted state (AFM mode). (f) By twisting $M_A$ and $M_B$ recurrently, the magnon chirality can be modulated and read out in form of the spin pumping voltage.

To probe spin pumping solely from $M_A$ or $M_B$, we synthesized an artificial AFM by spatially separating $M_A$ and $M_B$ into antiferromagnetically coupled $M_A/M_B$ multilayers so that spin pumping from the $M_A$ layer could be selected explicitly by growing the spin-current receiving layer (HM) next to the $M_A$ layer. Then, the spin pumping of the right-handed or left-handed $M_A$ precession can be specified in a quantitative manner [Fig. 1(a) and (b)].

Figure 2(a) illustrates Py(2.5)/Gd(3)/Py(2.5)/Gd(3)/Py(2.5)/Pt(6) multilayer (in nm), hereafter simplified as the Py/Gd multilayer. Different Curie temperatures of Py and Gd (for bulk materials, $T_C^{Py}$ = 872 K and $T_C^{Gd}$ = 293 K) accompanied by strong interfacial AFM coupling result in a compensation temperature $T_M$ [11]. At $T = T_M$, the magnetic moments of Py ($M_{Py}$) and Gd ($M_{Gd}$) are fully compensated. For $T > T_M$, we could achieve a Py-aligned phase ($M_{Py} > M_{Gd}$) with $M_{Py}$ parallel to $H$. For $T < T_M$, the Gd-aligned phase ($M_{Py} < M_{Gd}$) with $M_{Py}$ opposite to $H$ is accessible. With respect to the strong AFM coupling at the Py/Gd interface, the relatively weak ferromagnetic exchange in the Gd layer will result in a transition to a canted magnetic state (twisted state) in the presence of a sufficiently strong magnetic field [12]. Thus, a rich magnetic phase diagram can be achieved depending on $H$ and temperature [13,14].

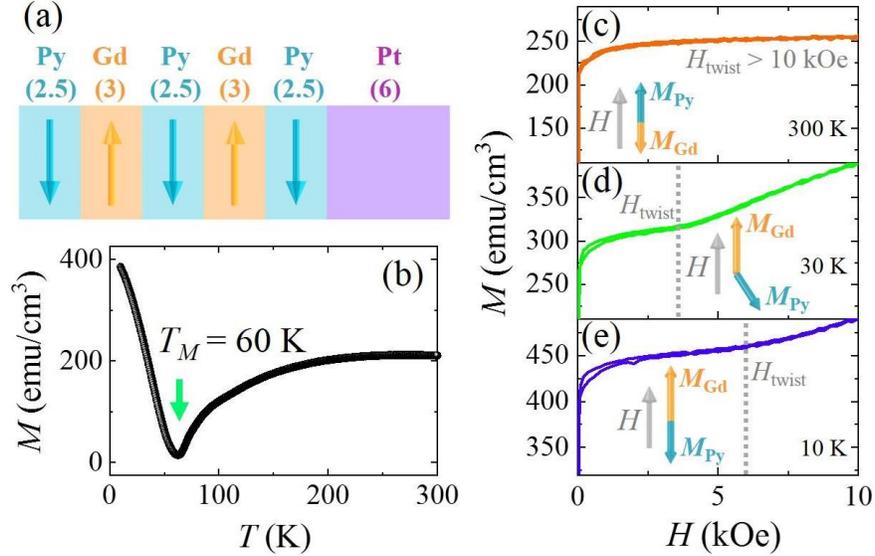

Fig. 2. Static magnetization of Py/Gd multilayer. (a) Sketch of the Py/Gd/Py/Gd/Py/Pt sample (the numbers in parentheses are thicknesses in units of nanometers). (b) Temperature dependence of in-plane magnetization at $H$ = 50 Oe; the green arrow indicates the compensation temperature $T_M$. The positive half branches of the hysteresis loops at (c) $T$ = 300 K, (d) $T$ = 30 K and (e) $T$ = 10 K. The twisted state is achieved at $H > H_{twist}$ (grey dotted lines in (d) and (e)), while the Gd-aligned phase is retained at $H < H_{twist}$ at $T$ = 30 K and 10 K. At $T$ = 300 K, only the Py-aligned phase is achievable within the $H$ range in our experiments. The Py-aligned phase, twisted state and Gd-aligned phase are illustrated in (c), (d) and (e). The perfect antiparallel alignment of $M_{Py}$ and $M_{Gd}$ is ensured in the Py-aligned phase and Gd-aligned phase.

Figure 2(b) depicts the temperature dependence of the in-plane magnetization at 50 Oe with a local minimum at $T$ = 60 K, revealing that $T_M$ = 60 K for the Py/Gd multilayer. Figure 2(c), (d) and (e) exhibit the positive half branches of the hysteresis loops at $T$ = 300 K, 30 K and 10 K, respectively. A nonlinear rise in the magnetization with $H$ was observed at $T$ = 30 K and 10 K, indicating the initiation of the twisted state at the critical field $H_{twist}$ (grey dotted lines in Fig. 2(d) and (e)) [14]. The Gd-aligned phase is retained when $H < H_{twist}$, and the twisted state could be achieved when $H > H_{twist}$ at $T$ = 30 K and 10 K. In contrast, $H_{twist}$ exceeds 10,000 Oe at $T$ = 300 K; thus only the Py-aligned phase is achievable within the $H$ range of our facilities at $T$ = 300 K. Additionally, the Py-aligned and Gd-aligned phases can also be revealed via the measurements of anomalous Hall effect [15]. Taking $M_{Py}$ and $M_{Gd}$ as $\mathbf{M}_A$ and $\mathbf{M}_B$ in Fig. 1, the chirality of $M_{Py}$ precession is controllable via the switching between Py-aligned phase and Gd-aligned phase, as well as the twisting of $M_{Py}$ and $M_{Gd}$ into the twisted state.

Next, we intend to manipulate the chirality of $M_{Py}$ precession in FMR mode via the switching between Py-aligned phase ($T = 300$ K) and Gd-aligned phase ($T = 10$ K), and conduct spin pumping measurements [16]. The experimental geometry is expressed in the azimuthal directions of in-plane $H$ ($\theta_H$) and $M_{Py}$ equilibrium ($\theta_{Py}$). $M_{Py}$ is aligned along the +**z** direction ($\theta_{Py} = 0°$) by an in-plane $H$ with $\theta_H = 0°$ at $T = 300$ K and $\theta_H = 180°$ at $T = 10$ K [Fig. 3(b) and (c)]. At $T = 300$ K, the master $M_{Py}$ causes right-handed $M_{Py}$ precession about the +**z** direction [inset in Fig. 3(h)]. At $T = 10$ K, the right-handed $M_{Gd}$ precession about the -**z** direction forces $M_{Py}$ to precess with left-handed chirality about the +**z** direction [inset in Fig. 3(i)]. Thus, we accomplish the right-handed $M_{Py}$ precession at $T = 300$ K and the left-handed $M_{Py}$ precession at $T = 10$ K with the same $M_{Py}$ equilibrium direction along the +**z** direction.

When spin pumping occurs in the Py/Gd multilayer, only the outermost Py layer (next to Pt) gives rise to the spin mixing conductance [17] at the Py/Pt interface. The inner Gd and Py layers are physically separated from the Pt layer and make null contributions to the spin mixing conductance at the Py/Pt interface, in light of the spin current penetration depth (~1 nm) in Py [18,19] and the negligible contributions of Gd [9,20]. This fact is further evidenced by a negligible spin mixing conductance at the Gd/Pt interface of control samples [15]. Hence, we can probe the spin pumping voltage $V(H)$ of the specific magnetic sublattice (the outermost $M_{Py}$), with either right-handed or left-handed chirality of $M_{Py}$ precession.

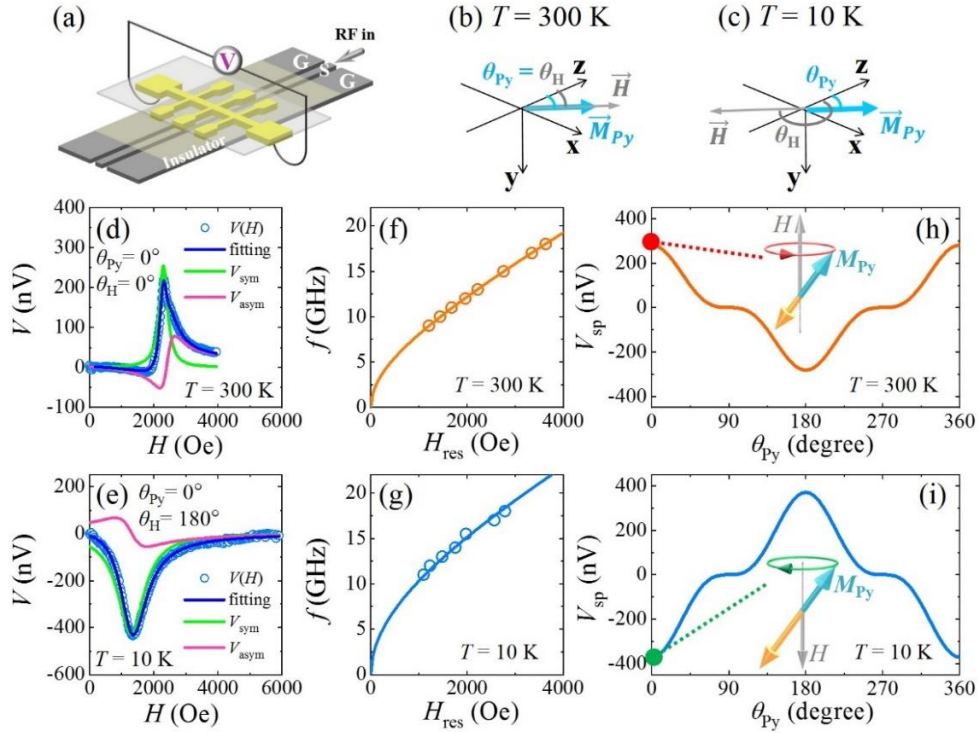

Fig. 3. Chirality-dependent spin pumping in the FMR mode. (a) Illustration of spin pumping measurements. The experimental geometry is characterized by the azimuthal directions of $H$ ($\theta_H$) and $M_{Py}$ equilibrium ($\theta_{Py}$). (b) $M_{Py}$ is parallel to $H$ at $T = 300$ K. (c) $M_{Py}$ is antiparallel to $H$ at $T = 10$ K. $V(H)$ signals ($f = 13$ GHz) are plotted with the fitting curves for (d) $\theta_{Py} = 0°$ and $\theta_H = 0°$ at $T = 300$ K and (e) $\theta_{Py} = 0°$ and $\theta_H = 180°$ at $T = 10$ K. The $H_{res}$ dependences of frequency $f$ (f) at $T = 300$ K and (g) at $T =$

10 K are effectively described by the Kittel equation, confirming the perfectly antiparallel alignment of $M_{Py}$ and $M_{Gd}$ in FMR. $\theta_{Py}$-dependent $V_{sp}$ was obtained by fitting at (h) $T = 300$ K and (i) $T = 10$ K. A positive $V_{sp}$ was produced by the right-handed $M_{Py}$ precessions ($\theta_{Py} = 0°$ and $\theta_H = 0°$, marked by the red dot in (h)). The negative $V_{sp}$ is produced by the left-handed $M_{Py}$ precession ($\theta_{Py} = 0°$ and $\theta_H = 180°$, marked by the green dot in (i)).

Figure 3(d) and (e) plot the $V(H)$ signals at $T = 300$ K and $T = 10$ K, respectively. A negative $V(H)$ is observed at $T = 10$ K with respect to the positive $V(H)$ at $T = 300$ K. At both temperatures, the dispersion relations between frequency $f$ and resonance field $H_{res}$ are effectively described by the Kittel equation [21]. Thus, we confirm that $V(H)$ originates from the FMR mode of the Py/Gd multilayer [21]. $M_{Py}$ and $M_{Gd}$ are perfectly antiparallel to each other during the precessions, leading to the right-handed $M_{Py}$ precession at $T = 300$ K and the left-handed $M_{Py}$ precession at $T = 10$ K. The right-handed (left-handed) $M_{Py}$ precession produces the positive (negative) $V(H)$. To exclude the contributions of the spin rectification effect (SRE) [22] and quantitatively determine the spin pumping voltage $V_{sp}$ due to pure spin current, we performed angular-dependent measurements of $V(H)$ [15,23,24]. The quantitative results reveal that $V(H)$ is mainly attributed to the spin pumping voltage $V_{sp}$. As plotted in Fig. 3(h) and (i), $V_{sp}$ shows positive polarity for right-handed chirality at $T = 300$ K (red dot at $\theta_{Py} = 0°$ in Fig. 3(h)) in comparison to the negative polarity for left-handed chirality at $T = 10$ K (green dot at $\theta_{Py} = 0°$ in Fig. 3(i)). This result unambiguously demonstrates that the spin polarization of the spin current is determined by the chirality of the spin precession rather than the spin equilibrium direction. $V_{sp}$ is a good measure of chirality for spin precession.

Spin Hall angle of Pt retains the same sign in the temperature range of our measurements and makes no impact on our conclusion [25,26]. The normalized $V_{sp}$ signal is linearly proportional to $f$ [15], in accordance with the spin pumping theory [17,27]. Hence, the pure spin current from coherent spin pumping is confirmed to be the origin of $V_{sp}$. Thermal voltage (incoherent pumping) is not evidenced in our experiments [28].

It is worth noting that $V_{asym}$ due to SRE reverses the sign when switching the chirality of $M_{Py}$ precession, as shown in Fig. 3(d) and (e). According to the quantitative measurements [15], $V_{asym}$ is mainly attributed to anisotropic magnetoresistance (AMR). In the scenario of AMR-related SRE [22,29], the radio frequency (*rf*) current $I(t) = I_{rf}\cos(2\pi f \cdot t)$ and the oscillating resistance $R(t) = R_0 - \Delta R_{AMR}sin^2[\theta_H + \Delta\theta(t)]$ are taken into account, where $\Delta R_{AMR}$ is the magnitude of AMR and $\Delta\theta(t)$ is the time dependent cone angle of magnetization precession. The product of $I(t)$ and $R(t)$ causes a DC $V_{asym}$ which is proportional to $cos(2\theta_H)cos(\theta_H)\sin\Phi$, where $\Phi$ is the relative phase between $I(t)$ and $\Delta\theta(t)$ [23]. The opposite chirality of magnetization precession would cause the opposite $\Phi$, leading to the opposite sign of $V_{asym}$. Hence, $V_{asym}$ of SRE is also a good measure of chirality. To the best of our knowledge, it is the first time to propose a measure of magnon chirality via SRE. Spin-current receiving layer (HM) is unnecessary in SRE measurement, which would be a salient advantage in the specific applications [30,31].

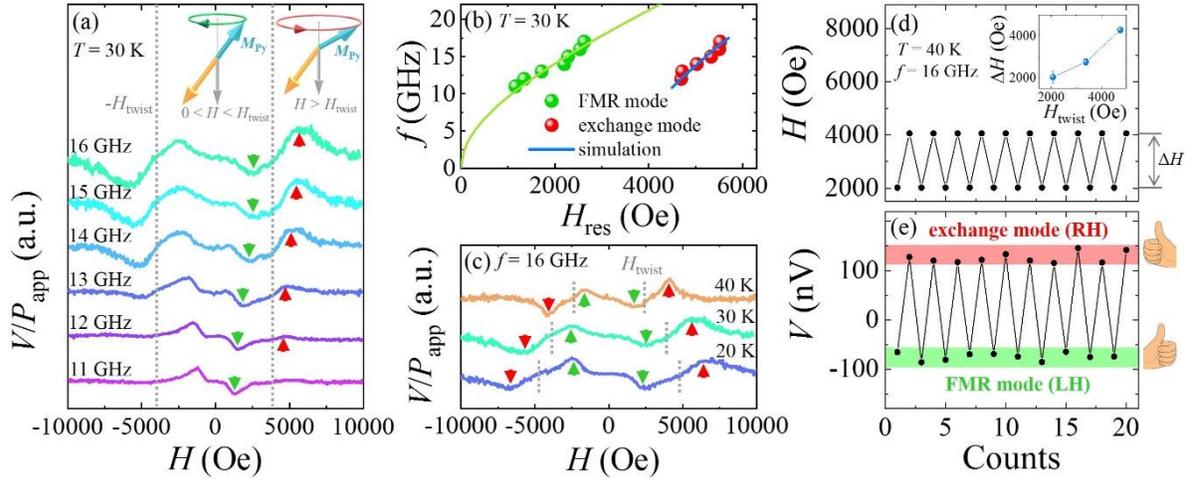

Fig. 4. FMR mode with left-handed (LH) chirality and exchange mode with right-handed (RH) chirality. (a) Normalized $V/P_{app}$ signals at $T = 30$ K. The FMR mode is observed when $H < H_{twist}$, and the exchange mode is observed when $H > H_{twist}$. (b) The dispersion relations between $f$ and $H_{res}$ of the FMR mode and exchange mode at $T = 30$ K. The FMR mode is confirmed through fitting by the Kittel equation. The $H_{res}$ dependence of $f$ for the exchange mode is calculated by the micromagnetic simulation. (c) Normalized $V/P_{app}$ signals ($f = 16$ GHz) at 20 K, 30 K and 40 K. FMR mode (exchange mode) is indicated by green (red) arrows. Grey dotted lines indicate the strength of $H_{twist}$. (d) The recurrent switching of the external field $H$ can modulate (e) the chirality as well as the $V(H)$ signal. The change of $H$ ($\Delta H$) for this modulation can be reduced by decreasing $H_{twist}$, as shown in the inset of (d).

Subsequently, we intend to manipulate the chirality of $M_{Py}$ precession via the twisting of $M_{Py}$ and $M_{Gd}$. Spin pumping measurements were performed at $T = 30$ K in the same experimental geometry of $T = 10$ K ($\theta_{Py} = 0°$ and $\theta_H = 180°$). As shown in Fig. 4(a), negative $V(H)$ values were observed when $H < H_{twist}$ (green arrows), which corresponded to the left-handed $M_{Py}$ precession of the FMR mode (confirmed via fitting by the Kittel equation in Fig. 4(b)). A positive $V(H)$ (red arrows) emerges in the twisted state ($H > H_{twist}$) when $f$ exceeds 12 GHz, corresponding to the second resonance mode [14]. This resonance mode is the exchange mode arising from the twisting of $M_{Py}$ and $M_{Gd}$ in the twisted state. The positive $V(H)$ of this mode indicates the right-handed $M_{Py}$ precession, which is confirmed via micromagnetic simulation [15,32]. The coexistence of the FMR and exchange modes with opposite chiralities was observed at a series of temperatures (Fig. 4(c)). As $H_{twist}$ (grey dotted lines in Fig. 4(c)) and $H_{res}$ of the exchange mode (red arrow in Fig. 4(c)) decline simultaneously with increasing temperature, the correlation between the exchange mode and twisted state is unambiguously clarified [inset in Fig. 4(d)]. It has been verified that the $V(H)$ of the FMR mode is mainly attributed to $V_{sp}$. We also examine the SRE signals of the exchange mode, and $V(H)$ of the exchange mode is mainly attributed to $V_{sp}$ [15]. Overall, the left-handed $M_{Py}$ precession of the FMR mode produces a negative $V_{sp}$, and the right-handed $M_{Py}$ precession of the exchange mode produces a positive $V_{sp}$.

The coexistence of the FMR mode (FM feature) and exchange mode (AFM characteristic) provides the opportunity to switch the chiralities by tuning $H$ at the fixed frequency and temperature. Figure 4(d) and (e) depict the modulation of chiralities by switching $H$ recurrently between 4 kOe and 2 kOe at $T = 40$ K and $f = 16$ GHz. The right-handed chirality (exchange mode) and left-handed chirality (FMR mode) were recurrently triggered and electrically read out by measuring $V(H)$ signals, i.e., information encoded in form of the chiralities can be modulated and read out. This result explicitly illustrates a prospective application of the artificial AFM in spintronics using chiral magnons as information carriers. The right-handed magnons could be treated as "spin up" in addition to the left-handed magnons being treated as "spin down" in the operations [33]. The separation between two modes ($\Delta H$) can be further reduced by decreasing $H_{twist}$ so that a smaller change in $H$ would be needed [inset in Fig. 4(d)]. This discovery promises artificial AFMs a great advantage over natural AFMs due to the ease of manipulating the magnon chirality. The coexistence of the FMR and exchange modes is also expected when $T > T_M$, which was realized in the Fe/Gd multilayer even at room temperature (an ideal working temperature for spintronics devices). We presented the detailed result in Supplementary S8 [15]. The aforementioned spin pumping measurements were performed in magnon mode with wave vector $k = 0$, which should be valid when $k \neq 0$. The right-handed and left-handed magnons exhibit opposite angular momenta, which is independent of the wave vector [34].

In summary, spin pumping was demonstrated in Py/Gd/Py/Gd/Py/Pt multilayers. The modulation and electrical readout of the chiralities were demonstrated in the artificial AFM, which was accessible even at room temperature. Our result opens the door for the prospective applications of chiral magnons in antiferromagnetic spintronics.

The authors thank Ran Cheng, Chanyong Hwang and Jing Shi for valuable discussions and assistance with the data analysis. This work is supported by the National Key Research and Development Program of China (Nos. 2016YFA0300804 & 2017YFA0303303), the National Natural Science Foundation of China (Nos. 11874072 & 11874416), the Strategic Priority Research Program of the Chinese Academy of Sciences (Grant No. XDB33000000), and the National Key Basic Research Project of China (Grant No. 2016YFA0300600).